\documentstyle[12pt]{article}
\begin{document}

\begin{titlepage}

\title{Parameter restrictions in a non-commutative
geometry model
do not survive standard quantum corrections}
\vspace{1.2cm}

\author{E.\' Alvarez, J.M. Gracia-Bond\'\i a  \thanks{and
Departamento
de F\'\i sica Te\'orica, Universidad de Zaragoza,50009
Zaragoza, Spain} and C.P.Mart\'\i n\\
Departamento de F\'\i sica Te\'orica \\Universidad
Aut\'onoma, 28049
Madrid, Spain}

\maketitle

\begin{abstract}
 We have investigated the standard one-loop quantum
corrections for a particularly simple
non-commutative geometry model containing fermions
interacting with a unique abelian
gauge field and a unique scalar through Yukawa couplings.
In this model
 there are certain relations among the different coupling
constants
quite similar to the ones appearing in the Connes-Lott
version of the standard model.
We find that it is not possible to implement those
relations in a renormalization-group
invariant way.

\vfill
\end{abstract}
\vspace{11cm}

\end{titlepage}

\newpage

\section{Introduction}
\vskip 1pc
There seems to be a growing consensus that non-commutative
geometry (NCG from now on)
(cf.\cite{connes,peligroamarillo}  for general reviews)is
one of the most important developments in mathematics
in the recent years.  In addition to that, Connes and Lott
(\cite{conneslott})
have invented a mechanism somewhat similar to the old
Kaluza-Klein idea
(but using discrete internal spaces instead, in such a way
that the Higgs field
is interpreted as a sort of gauge field in the internal
direction), to show that the standard $SU(3)
\times SU(2) \times U(1)$ model of strong and electroweak
interactions (SM in the sequel)
 appears naturally in this
framework. The fascinating point is that, when interpreted
in this way, not all the parameters of the
SM are free, but have to obey certain restrictions. In
particular,
the Higgs mass in terms of the top mass (neglecting all
other fermion masses) is
given by:
\begin{equation}
M_{H} = c(x) m_t
\end{equation}
Where $c(x)$ is a constant , depending on the parameter
$x$, which
 measures the splitting of the trace
between the leptonic and the quark sector ($-1 \leq x \leq
1$).
Actually, if the total Hilbert space of the fermions in
the Connes-Lott derivation
is $H=H_{leptons}\bigoplus H_{quarks}$, then the total NCG
Yang-Mills functional
is a convex combination:
\begin{equation}
\frac{1 + x}{2 }(YM)_{leptons} \bigoplus \frac{1-x}{2}
(YM)_{quarks}
\end{equation}

To be specific\cite{kastler},
\begin{equation}
c(x)^2 = 3 - \frac{9x^2 - 24 x + 15}{10 x^2 - 34 x + 28}\
\end{equation}
\par
As far as one can tell with our limited knowledge of
quantum field theory,
the only consistent way of imposing constraints among
different coupling
constants in a given model, is in a renormalization group
invariant way.
This is almost always a consequence of the Ward identities
corresponding to
some symmetry of the underlying action principle, although
this is not always
neccessary, as in Zimmermann's examples of "reduction of
coupling constants"
(cf., for example,\cite{zimmermann}).
\par
We could, of course, impose those restrictions as defining
the "physical"
renormalization scheme, even if they do not hold for
general and
sensible renormalization schemes.
But this would be most "unnatural", from the quantum field
theory point of view,
and besides, we do not see a compelling reason in the
Connes-Lott construction
to do that.
\par
It is then obviously of great interest to investigate
whether those restrictions
are indeed first integrals of the renormalization-group
flow. Were that the case,
it would point out either to a "hidden" symmetry of the
standard model, overlooked
fntil now, or else to some reduction mechanism of the
Zimmermann type. In either case
it would have been most remarkable.
\par
Although our main interest lies in the SM, there are many
technical complications
(coming essentially from the chiral character of the
model), which make the computation of
the one-loop
renormalization group of the model a nontrivial matter.
We have decided, in view of that, to study first a toy
model, in which we have succeeded
to include
some relations among the parameters quite similar to the
ones appearing in the
Connes-Lott version of the SM.
\section{The Non-commutative geometry
model}
\vskip	1pc
Using suitable generalizations of the Connes-Lott
construction one can
obtain a limited type of lagrangians only. The fact that
the SM is among them
is already quite remarkable. The toy model we are going to
discuss for the
remaining of this paper is obtained by choosing as our
manifold the product space
of $Z_2$ and (euclidean) spacetime M, including a trivial
one-dimensional
vector bundle on each piece. This is actually the simplest
nontrivial (in the NCG
 sense)
and non-pathological model available (anomalous models
result, for example, when
one considers the set theoretical union of spacetime with
discrete sets of points).
\par
The basic tool in NCG is the K-cycle, which in the
commutative case is equivalent to having a
gauge field coupled to massless fermions. In order to
obtain Higgs fields, one needs product
K-cycles, as in our toy model, were we took the product of
Dirac's K-cycle with the K-cycle giving
the geometry of $Z_2$, which contains in embryonic form
both the Higgs and the Yukawa couplings.
Details can be obtained from \cite{peligroamarillo}; our
present model can be
actually obtained by putting $Z = W = \phi_2 = 0 $ and
$\phi_1^{*} = \phi_1$ in the
computation of the NCG version of the SM in Section 7 of
that paper.
\par
In the model there are $N$ species of fermions,
$\psi_i$,all with the
same $U(1)$ charge with respect to an abelian gauge field
$A_{\mu}$,
 and with
different Yukawa couplings $g_i$ with a single scalar
field, of mass $2 M$,
and self-coupling $\lambda$.
\par
The Lagrangian (after rotating to Minkowski space)
is:\footnote{Let us point out
that there is a mass non-degeneracy condition implying
that $N \geq 2$}
\begin{equation}
L = -1/4 F_{\mu \nu} F^{\mu \nu} +1/2 \partial_{\mu} \phi
\partial^{\mu}
\phi + 1/2 M^2 \phi^2 - \frac{\lambda}{24} \phi^4 +
\sum_{j=1}^{N} i \bar{\psi_j}
\gamma^{\mu}( \partial_{\mu} - e A_{\mu}) \psi_{j} -g_j
\bar{\psi_j} \psi_j
\phi \end{equation}

With regard to the analogy with the standard model, it
seems more adequate to
think of the above fermions as "leptons" instead of
"quarks", because color
enters NCG through Poincare duality, which is trivial for
the algebra
$C^2 \bigotimes C^{\infty}(M)$.
\par
In our model the restrictions on the allowed values of the
different coupling
constants arise just because our product K-cycle contains
information about the
fermion mass spectrum only \footnote{In particular, the
mass of the heaviest
fermion is interpreted as the inverse distance between the
two leaves of
spacetime}; the only remaining freedom being the scale of
the $U(1)$
connection; that is, the charge.
\par
To be specific, the parameters above must obey three
different relations among
themselves:
\begin{equation}
\sum_{i=1}^{N} g_{i}^2 = \frac{N e^2}{2}
\end{equation}
\begin{equation}
\lambda = 6 N e^2 \frac{tr m_{\perp}^4}{(tr m^2)^2}
\end{equation}
\begin{equation}
M^2 = \frac{tr m_{\perp}^4}{ tr m^2}
\end{equation}

The fermionic mass spectrum is determined by the diagonal
matrix $m$, and
$m^2_{\perp} = m^2 - (tr m^2)/N$, where the parameters
$m_i$ and the Yukawa couplings must obey the relation:
\begin{equation}
g_{i}^{2} = \frac{N e^2 m_{i}^{2}}{2 tr m^2}
\end{equation}

When one of the fermions (for example, the first) is much
heavier than all the
others, the second equation reduces to:
\begin{equation}
\lambda = 6 (N-1) e^2
\end{equation}
and the third one reduces in turn to
\begin{equation}
N M^2 = (N-1) m_{1}^2
\end{equation}
\par
The last two relations have close analogous in the
Connes-Lott version of the
SM(cf. \cite{conneslott,peligroamarillo})\footnote{There
is some
controversy on this point (\cite{kastler,coquereaux})}.
The first one
has an analogous also,
although apparently it has never been explicitily written
in the
literature.
 Let us write it here for completeness. In terms of the
parameter $x$ we introduced in the first paragraph, it
yields:
\begin{equation}
\frac{1 + x}{2} \sum_{i=leptons} g_{i}^2 + \frac{3(1 -
x)}{2} \sum_{i=quarks}
g_{i}^2
 = \frac{N(2 - x) e^2}{4 sin \theta_{W}^2}
\end{equation}

\section{The one-loop beta functions of the model}

Our toy model is obviously anomaly free and
renormalizable. Besides, it is
non-chiral, so that we can freely use dimensional
regularization without having
to worry about how to define $\gamma_5$. The one loop beta
functions in the MS
scheme are easily shown to be\footnote{To perform
perturbative
computations one has to take into account that in our
model $\langle
\phi\rangle  \neq 0$}:
\begin{equation}
\beta_{g_{i}} = \frac{1}{16\pi^2} (5 g_{i}^3 - 6 e^2 g_i +
2 g_i \sum_{j \neq i}
g_{j}^2)
\end{equation}

\begin{equation}
\beta_{e} = \frac{N}{12 \pi^2} e^3
\end{equation}

\begin{equation}
\beta_{m_{i}} = \frac{3 m_{i}}{16\pi^2} (- 2 e^2 +
g_{i}^2)
\end{equation}

\begin{equation}
\beta_{M^2} =\frac{ M^2}{16\pi^2} (\lambda + 4 \sum_i
g_{i}^2)
\end{equation}

\begin{equation}
\beta_{\lambda} = \frac{1}{16 \pi^2} (- 48 \sum_{i}
g_{i}^4 + 3 \lambda^2 + 8
\lambda \sum_{i} g_{i}^2)
\end{equation}

Let us make two technical comments here, for the sake of
completeness: first, the
formal Ward identities coming from both the discrete
spontaneously broken
symmetry and the $U(1)$ gauge symmetry of our model hold
authomatically in our
substraction scheme \footnote{Actually, the dimensionless
counterterms
of the unbroken phase are the same as their counterparts
in the broken
phase,\cite{collins}}. Secondly, equations (5) and (6)
yield $\lambda
\sim e^2$
and at least one Yukawa $g_j \sim e$; this means that
contributions of
the order $g_i \lambda^2$ and $M^2 \lambda^2$ constitute
higher order
corrections, and are thus neglected here.
\par
Now let us look at the physical implications of these
values for the beta
functions of the model.
\par
It is easily seen from the renormalization group equations
for a generic
coupling constant (including masses), $g$
\begin{equation}
\mu \frac{d g(\mu)}{d \mu}\ = \beta(g(\mu))
\end{equation}
that the conditions the $\beta$-functions of the model
ought to obey
in order for the NCG conditions be first integrals of the
above
differential
equations are:
\begin{equation}
2 \sum_{i} g_i \beta_{g_i} = N e \beta_e
\end{equation}
\begin{equation}
\beta_{\lambda} = \frac{12 N}{(tr m^2)^2}(tr
m_{\perp}^{4}(e \beta_e - 2
e^2 \sum_{i} \frac{m_i \beta_{m_i}}{tr m^2}) + 2 e^2
\sum_i m_i
\beta_{m_i}(m_{i}^2 - \frac{tr m^2}{N}))
\end{equation}
\begin{equation}
\beta_{M^2} = \sum_i \frac{2m_i \beta_{m_i}}{tr m^2} (-
\frac{tr
m_{\perp}^4}{tr m^2} + 2 m_i^2 - \frac{2 tr m^2}{N})
\end{equation}

The fact that these equations are not satisfied
identically in parameter
space means that if one imposes them at one scale $\mu_0$,
the flow
takes the system away from the constraint surface.

\section{Conclusions}
\vskip 1pc

What can we conclude from the preceding analysis?
One thing, at least, seems to be clear: there is no
underlying hidden symmetry
in our model, nor a Zimmermann-like mechanism of reduction
of coupling
constants, because in both cases the constraint surface
would remain invariant
under the renormalization group flow.
\par
One can always argue, however, that the only physically
admissible
renormalization scheme is the one preserving the
constraints coming from NCG;
which is certainly technically possible, because we have
less restrictions than
coupling constants.
\par
In a different, and perhaps deeper, vein, it can certainly
be maintained that the
structure of the non-commutative geometry underlying the
model requires a
drastic change in the standard quantization rules based
upon replacing Poisson
(or Dirac) brackets by commutators of operators acting in
some linear space.
\par
Our present work, being rooted in ordinary quantum
mechanics is unable to rule
out such a possibility.

\section{Acknowledgements}
We are grateful to A. Gonz\'alez-Arroyo and L.
Ib\'a\~{n}ez for enlightening
conversations. We also thank M.A.R. Osorio and M.A.
V\'azquez-Mozo for help with
some numerical computations. This work has been supported
by CICYT
(Spain).


\begin{thebibliography}{99}
\bibitem{connes} A.Connes,Pub. Math.
I.H.E.S.,$\underline{62}$(1985) 257
A.Connes,Geometrie non-conmutative
(Intereditions,Paris,1990)

\bibitem{peligroamarillo} J.C.Varilly and J.M.
Gracia-Bondia,preprint
FTUAM,29/92,DFTUZ 92-20 J.Geom. and Physics, to appear

\bibitem{conneslott}A.	Connes and J. Lott,Nucl.  Phys.
B. (Proc.
Suppl.),$\underline{18}$ (1990)29 A. Connes and J.
Lott,"The metric
aspect of non-commutative geometry", preprint (1992)

\bibitem{kastler}D. Kastler and T. Schvcker,Marseille
preprint CPT-2724
(1992)

\bibitem{zimmermann}W.	Zimmermann, Comm.  Math.
Phys.$\underline{97}$(1985),142

\bibitem{coquereaux}R. Coquereaux,Lectures at the Karpacz
School
CERN preprint, CERN-TH-6552-92 (1992)

\bibitem{collins} J.C.	Collins,"Renormalization"
(Cambridge University
Press)

\end{thebibliography}
\end{document}